\def\dbcol{double column single spacing}
\ifdefined\dbcol
\ifdefined\ACM
	\documentclass{sig-alternate-10pt}
	\makeatletter
	\def\@copyrightspace{\relax}
	\makeatother
\else
	\documentclass[conference,10pt,letterpaper]{IEEEtran}
	\IEEEoverridecommandlockouts
	\makeatletter
	\def\ps@headings{%
	\def\@oddhead{\mbox{}\scriptsize\rightmark \hfil \thepage}%
	\def\@evenhead{\scriptsize\thepage \hfil \leftmark\mbox{}}%
	\def\@oddfoot{}%
	\def\@evenfoot{}}
	\makeatother
\fi

\usepackage[left=.5in,right=.5in,top=.55in,bottom=.5in]{geometry}
\else
	\documentclass[conference,12pt,onecolumn]{IEEEtran}
\fi

\ifdefined\COMSOC
  \usepackage[nocompress]{cite}
\else
  \usepackage{cite} 
\fi
\ifdefined\ACM 
	\usepackage{amsmath}
\else
	\usepackage[cmex10]{amsmath} 
	\usepackage{amsthm} 
	\usepackage[pdftex]{graphicx} 
\fi

\ifdefined\META     \usepackage{stmaryrd}    \fi 

\newcommand{\begproof}{\ifdefined\dbcol\begin{IEEEproof}\else\begin{proof}\fi}
\newcommand{\Endproof}{\ifdefined\dbcol\end{IEEEproof}\else\end{proof}\fi}
\newcommand{\metacom}[1]{\ifdefined\META\bluepure{$\blacktriangleright$}#1\bluepure{$\rrbracket$}\fi}
\newcommand{\metafoot}[1]{\ifdefined\META\footnote{\bluepure{$\blacktriangleright$}#1}\fi}

\usepackage[font=small]{subfig} 

\usepackage{hyperref} 

\usepackage{amssymb}
\usepackage{bbm} 
\usepackage{verbatim}
\usepackage{color}
\usepackage[normalem]{ulem}
\usepackage[ruled,lined,linesnumbered]{algorithm2e}

\newtheorem{thm}{Theorem}
\newtheorem{lem}{Lemma}
\newtheorem{prop}{Proposition}

\newcommand{\fref}[1]{Fig.~\ref{#1}}
\newcommand{\tref}[1]{Table~\ref{#1}}
\newcommand{\sref}[1]{Section~\ref{#1}}
\newcommand{\thmref}[1]{Theorem~\ref{#1}}
\newcommand{\lref}[1]{Lemma~\ref{#1}}
\newcommand{\pref}[1]{Proposition~\ref{#1}}
\newcommand{\cref}[1]{Corollary~\ref{#1}}

\newcommand{\aref}[1]{Algorithm~\ref{#1}}

\newcommand{\vect}[1]{\boldsymbol{#1}}

\newcommand{\bluepure}[1]{{\color{blue}{#1}}}

\newcommand{\eps}{\epsilon}
\newcommand{\ovl}{\overline}
\newcommand{\nn}{\nonumber}
\newcommand{\opd}{\operatorname{d}\!}

\newcommand{\inv}[1]{\frac{1}{#1}} 
\DeclareMathOperator*{\argmax}{arg\,max}

\usepackage{enumitem}

\usepackage{caption}
\usepackage{fancyhdr}
\fancypagestyle{empty}{
  \fancyhf{}
  \fancyhead[C]{\footnotesize{T. Luo, S. S. Kanhere, H-P. Tan, F. Wu, and H. Wu, ``Crowdsourcing with Tullock contests: A new perspective'', {\it Proc. IEEE INFOCOM}, 2015, pp. 2515-2523.}}
}

\usepackage{datetime}
\title{Crowdsourcing with Tullock Contests:\\A New Perspective}
\author{
  \IEEEauthorblockN{Tie Luo\IEEEauthorrefmark{1}, Salil S. Kanhere\IEEEauthorrefmark{2}, Hwee-Pink Tan\IEEEauthorrefmark{1}, Fan Wu\IEEEauthorrefmark{3}, Hongyi Wu\IEEEauthorrefmark{4}
  \IEEEauthorblockA{\small 
	\IEEEauthorrefmark{1}Institute for Infocomm Research, A*STAR, Singapore\\
	\IEEEauthorrefmark{2}School of Computer Science and Engineering, The University of New South Wales, Australia\\
	\IEEEauthorrefmark{3}Shanghai Key Laboratory of Scalable Computing and Systems, Shanghai Jiao Tong University, China\\
	\IEEEauthorrefmark{4}Center for Advanced Computer Studies, University of Louisiana at Lafayette, USA\\
    E-mail: luot@i2r.a-star.edu.sg, salilk@unsw.edu.au, hptan@i2r.a-star.edu.sg, fwu@cs.sjtu.edu.cn, wu@cacs.louisiana.edu}
}}

\begin{document}
\maketitle
\thispagestyle{empty}

\begin{abstract}
Incentive mechanisms for crowdsourcing have been extensively studied under the framework of all-pay auctions. Along a distinct line, this paper proposes to use Tullock contests as an alternative tool to design incentive mechanisms for crowdsourcing. We are inspired by the conduciveness of Tullock contests to attracting user entry (yet not necessarily a higher revenue) in other domains. In this paper, we explore a new dimension in {\em optimal Tullock contest design}, by superseding the contest prize---which is {\em fixed} in conventional Tullock contests---with a {\em prize function} that is dependent on the (unknown) winner's contribution, in order to maximize the crowdsourcer's utility.  We show that this approach leads to attractive practical advantages: (a) it is well-suited for rapid prototyping in fully distributed web agents and smartphone apps; (b) it overcomes the {\em disincentive} to participate caused by players' antagonism to an increasing number of rivals. 
Furthermore, we optimize conventional, fixed-prize Tullock contests to construct the {\em most superior} benchmark to compare against our mechanism. Through extensive evaluations, we show that our mechanism significantly outperforms the optimal benchmark, by over three folds on the crowdsourcer's utility cum profit and up to nine folds on the players' social welfare.
\end{abstract}

\section{Introduction}\label{sec:intro}

Crowdsourcing represents a new problem-solving model that elicits solutions, ideas, data, etc.---referred to as {\em contributions}---from an undefined, generally large group of people. Classic examples include Amazon Mechanic Turk, Yahoo! Answers, GalaxyZoo.org, TopCoder.com, etc.
Recently, a new variant of crowdsourcing called participatory sensing emerged as a new data-collection model, which elicits sensor data contributed from user-owned mobile devices such as smartphones. Examples include GreenGPS\cite{greengps10mobisys},
LiveCompare\cite{livecomp09},
ContriSense:Bus\cite{lau11ucc} and Waze.com, to name a few.

Pivotal to the viability of all such crowdsourcing systems, is whether there is enough {\em incentive} to attract sufficient participation. A large body of prior work \cite{yang12mobicom,kout13infocom,luo14infocom,DV09csallpay,AS09ICIS,SODA12,luo14mass,luo12secon} 
has been dedicated to designing incentive mechanisms for such scenarios, where each incentive mechanism essentially determines some reward according to users' contributions.
The commonly adopted approach turns out to be {\em auctions}, where each bidder tenders a bid (e.g., planned sensing duration\cite{yang12mobicom} or desired payment\cite{kout13infocom}) to the crowdsourcer, who will then choose the highest or lowest bidder(s) as the winner(s) to give out some reward. A widely used form among these auctions is {\em all-pay auctions}\cite{DV09csallpay,AS09ICIS,SODA12,luo14infocom,luo14mass}, which nicely captures the scenario where each bid represents some {\em irreversible} effort. In other words, effort has to be {\em sunk} at the time of bidding, for example working out a solution to a problem, or sensing and sending data through a smartphone.

A distinctive characteristic of auctions in general, is that they are {\em perfectly discriminating}\cite{Hillman89}: the best (highest or lowest) bidder wins with probability one while the others lose for sure. Thus in all-pay auctions, due to the inevitable sunk cost (every bidder has to pay for his bid regardless of whether he wins the auction or not), all bidders substantially {\em shade} (i.e., decrease) their bids for fear of loss\cite{Krishna09}.  Furthermore, if a bidder believes that there exists some other bidder who will bid higher than him, he will choose not to bid at all.  Indeed, as Franke et al. \cite{Franke14} pointed out, all-pay auctions are so discriminative that, under a complete-information setting, only two (strongest) players will enter the auction in equilibria and only one of them will have a positive expected payoff.
Clearly, this is not desirable in many crowdsourcing campaigns that favor a large participant pool (yet not necessarily higher revenue) for the sake of more population diversity (e.g., LiveCompare\cite{livecomp09}) and/or larger geographic coverage (e.g., GreenGPS\cite{greengps10mobisys} and Waze).

Taking a radically different approach, this paper proposes to use {\em Tullock contests} as an alternative framework to design incentive mechanisms for crowdsourcing. Fathered by Tullock's seminal work \cite{Tullock80}, Tullock contests represent a distinct contest regime that is {\em imperfectly discriminating}: every player has a strictly positive probability to win (determined by a contest success function) as long as he bids.\footnote{In an asymptotic limiting case, Tullock contests subsume all-pay auctions. However, they are generally classified as two different contest regimes.} This characteristic makes Tullock contests highly conducive to attracting user entry, especially weak players\cite{Franke14},\footnote{A ``strong'' or ``weak'' player in this paper refers to a player with a strong or weak {\em type}; type, as a term in Bayesian games and mechanism design, refers to the private information held by a player, such as his valuation of the auctioned item or his production cost. Therefore, a strong player means a player with high valuation or low cost, and vice versa.} which is of high practical interest because weak players often constitute the majority of potential participants in crowdsourcing. This explains why, in reality, we see {\em lotteries}---the simplest form of Tullock contests---much more often, and usually engaging a larger number of participants, than all-pay auctions.

On the other hand, Tullock contests are not necessarily superior to all-pay auctions in terms of {\em revenue}, or total user contribution. In fact, there is no conclusive theory as to which contest regime revenue-dominates the other in general\cite{allpay-lot-wp13,Fang02lot,Franke14}, and an experimental study also shows that revenue in all-pay auctions may be independent of the number of participants at some stable state\cite{allpay06exp}. Intuitively, the reason is that (a) the fierce competition induced by all-pay auctions efficaciously incentivizes (a small number of) strong players to exert high effort, while (b) the mild competition in Tullock contests attracts a medium amount of contributions from more (albeit weaker) players. Therefore, all-pay auctions are advantageous in eliciting the highest-quality contribution from the strongest players, such as selecting the best performer for a competition, while Tullock contests are superior in attracting more users and hence are beneficial to population diversity and geographic coverage, such as in lifestyle\cite{livecomp09} and transport mobile apps\cite{greengps10mobisys,lau11ucc}.

Thus Tullock contests are complementary to all-pay auctions. We note that these two frameworks have been compared in terms of their respective benefits in other domains such as fundraising\cite{allpay-lot-wp13}, lobbying\cite{Fang02lot}, and general contests\cite{lottery13apa}. We find that the comparison results therein can apply to crowdsourcing in principle. Therefore, the rest of the paper will focus, within the regime of Tullock contests, on {\em optimizing} this framework for crowdsourcing.

The objective, as is most common, is to maximize the crowdsourcer's revenue. To this end, we explore a new dimension in the space of Tullock contest design, by superseding the contest prize---which is {\em fixed} in conventional Tullock contests---with a {\em prize function} that is dependent on the (unknown) winner's contribution.  The rationale is to create a {\em two-tier incentive} to improve the efficacy of Tullock contests: the first tier, as exists in conventional contests as well, is for a player to win a prize by competing with and outdoing other players; on top of this, the second-tier incentive is for each player to {\em outdo himself} in order to {\em amplify the prize}. Logically, this approach also leads to a change of the crowdsourcer's objective: maximizing revenue becomes maximizing {\em profit}---revenue minus the (non-fixed) cost (prize)---which is also his utility. To the best of our knowledge, this paper is the first that introduces prize as a function into {\em optimal Tullock contest design}, a subject that is being pursued since the 1990's\cite{optlot91} following Tullock's seminal work\cite{Tullock80} in 1980.

Ultimately, the ``new perspective'' in the title of this paper has dual interpretations: (a) a new alternative mechanism-design framework for crowdsourcing, and (b) a novel dimension of optimal Tullock contest design.

To find an appropriate benchmark for a new mechanism designed as such to compare against, we need a fixed-prize Tullock contest. However, even this conventional and seemingly simple case turns out to be challenging---a general analytical solution to its equilibria does not exist and only numerical ones are available in the literature\cite{Fey08,Ryvkin10,Wasser13}.  Furthermore, we go one significant step beyond prior art, by not only {\em solving} equilibria of such conventional contests, but also {\em optimizing} the contests by finding the ``best'' equilibrium in terms of the same (utility-maximizing) objective. This allows us to compare our proposed mechanism with the best possible benchmark.

Extensive performance evaluations reveal that our mechanism outstrips the optimal benchmark by a remarkable margin: a 250\% increase in the crowdsourcer's utility (profit) and a 830\% improvement in the players' aggregate utility (social welfare). The improved performance achieved with these two typically {\em competing} metrics reflects a highly desirable ``win-win'' situation.

\subsection{Related Work}

Even in their simplest form, Tullock contests are analytically more challenging to tackle than most classic auctions. This is particularly true in the incomplete-information setting\footnote{Briefly speaking, with complete information all the players are informed of all the others' types, while with incomplete information each player only knows his own type. \sref{sec:model} explains this in more detail.} which is a more realistic setting for crowdsourcing. Specifically, the equilibria of most classic auctions with complete information, or with incomplete information and symmetric players, can be solved in closed form; but Tullock contests with incomplete information is generally intractable in analytical means, even in the simplest form (lottery)\cite{Konrad09book}. This can be attributed to the {\em double uncertainty}: in auctions with incomplete information, the uncertainty about other players' types is the only source of uncertainty; but in Tullock contests, the imperfectly discriminating nature---or more specifically the probabilistic winner selection (unlike in auctions the highest bid guarantees winning)---creates another source of uncertainty.

As a consequence, the literature on Tullock contests exclusively deals with the complete-information setting or restrictive versions of the incomplete-information setting (e.g., only two discrete types for two players\cite{MY04}). It was not until 2008 that Fey\cite{Fey08} first proved the existence of a (symmetric) equilibrium for a lottery with incomplete information. However, the model is limited to two players and uniform distribution, and the equilibrium strategy is only numerically characterized.

A subsequent breakthrough was made by Ryvkin\cite{Ryvkin10}, who extended Fey's model\cite{Fey08} by allowing for more than two players, arbitrary continuous distributions, and a more general contest success function. He also proved the existence of equilibria (leaving uniqueness as future work), following the spirit of \cite{Fey08}. Still, the equilibrium strategy was only numerically computed, due to the limited analytical tractability of Tullock contests.  So far, the only known analytical solution to equilibria with a continuous type distribution, is due to Ewerhart\cite{Ewer10IPV} who constructed a rather special distribution to obtain a closed-form expression. However, the distribution is rather complex and not generalizable, and the model is still limited to a two-player lottery only.

In this paper, under a general crowdsourcing model with incomplete information, we derive the {\em optimal prize function} that maximizes the crowdsourcer's utility cum profit. Surprisingly, our solution of the unique Bayesian Nash equilibrium (a) can be expressed in a simple and {\em closed form} in general cases, and (b) is {\em agnostic} to the number of players. These are in stark contrast to prior art, and in practical terms, imply that our mechanism (a) can be easily implemented in web agents and smartphone apps that act in a fully distributed manner, and (b) overcomes the {\em disincentive} to participate caused by player's {\em antagonism} to an increasing number of rivals.

Along the line of optimal Tullock contest design, two general directions have been pursued in prior work. One stream of research explores whether the prize should be allocated to a single winner or all the players in a hierarchical manner. For example, \cite{optlot91} applies a rank-dependent expected utility model to a lottery in which the prize was divided into, according to ranks, a few large prizes and a large number of small prizes. However, \cite{optSymTullock12} proves that it is optimal to give the entire prize to a single winner in a symmetric equilibrium. The other direction focuses on whether and how to bias players in such a way that induces the maximum revenue. For instance, \cite{Franke14} allows the crowdsourcer to assign different weights (preferences) to players in a discriminative manner for revenue maximization. \cite{lottery13apa} proves that a biased lottery (like \cite{Franke14}) achieves the same revenue as a biased all-pay auction, when both are fully optimized. In the fair case, \cite{lottery13apa} proves that an optimal lottery is always superior to an optimal all-pay auction.

Our proposed approach represents a new dimension in the design space of Tullock contests. Provisioning contest prize as a function (of the unknown winner's contribution) sets this work apart from all prior work on Tullock contests in which prizes are fixed and known ex ante.

\subsection{Contributions}

The main contributions of this paper are summarized below:
\begin{enumerate}[leftmargin=1em]
\item This work is the first attempt in the crowdsourcing literature that uses Tullock contests as a new framework to design incentive mechanisms.
\item We explore a new dimension of optimal Tullock contest design by provisioning the prize as a function. We demonstrate the simplicity of our approach which makes it particularly well-suited for rapid prototyping in fully distributed web agents and smartphone apps. We also show that our approach overcomes the disincentive caused by players' antagonism to an increasing number of rivals.
\item As a byproduct of this work, we construct an optimal fixed-prize Tullock contest as the benchmark for comparison and outline a step-by-step algorithm for it. This benchmark precisely falls in line with standard Tullock contests on which extensive studies are based. Therefore, the benchmark, its constructing algorithm and the associated performance analysis, are highly instructive for future research on Tullock contests. 
\item Our last contribution, which is {\em not} mentioned above, is that we introduce a new parameter---the crowdsourcer's valuation of user contribution---into the contest model, and show that it has an {\em exponential} positive effect on the performance of both our and conventional mechanisms. In practical terms, this means that a crowdsourcer can accrue higher payoff by improving his business processes via a better utilization of the crowdsourced contributions.
\end{enumerate}

The rest of this paper proceeds as follows. \sref{sec:model} presents our model with our proposed mechanism, and \sref{sec:analysis} analyzes the model to derive the optimal Tullock contest. The optimal benchmark is then constructed in \sref{sec:opt-fixed}. Following that, an extensive performance evaluation is provided in \sref{sec:numeric} which demonstrates key results and offers intuition as well. Finally, \sref{sec:conc} concludes.

\section{Contest Model}\label{sec:model}

Sitting at the core of a Tullock contest framework is a {\em contest success function} (CSF) which specifies the probability that a player $i=1,2,...,n$ who exerts (or ``bids'') effort $b_i$ wins the contest. Our model assumes a very general form of CSF:
\begin{align}\label{eq:csf}
\Pr(b_i) = \frac{g(b_i)}{\sum_{j=1}^n g(b_j)}
\end{align}
which generalizes the classic Tullock CSF, $b_i^r/\sum_{j=1}^n b_j^r$ where $r>0$, as well as the most-studied form of $r=1$ (also known as {\em lottery}). In \eqref{eq:csf}, $g(\cdot)$ is a nonnegative, strictly increasing function that satisfies $g(0)=0$ and converts player $i$'s effort $b_i$ into his {\em contribution} $\xi_i$.  For mathematical convenience, we assume that $g(\cdot)$ is twice differentiable and concave, which captures the common phenomenon of diminishing marginal return when exerting effort.  When $b_i=0$ for all $i$, i.e., no one exerts effort, we assume $\Pr(b_i)=0$, i.e., no one will win any prize.\footnote{\label{foot:nonzero}An alternative and more commonly adopted practice in the literature, is assuming $\Pr(b_i)=1/n$. However, the prizes therein are all fixed, and hence if $b_i=0$ for all $i$, a player $j$ will have incentive to deviate by exerting an infinitesimal effort $\eps>0$ to increase his payoff by $(1-\inv{n})v_j - O(\eps)$ where $v_j$ is his valuation of the prize. Therefore, an ``all-zero-bid'' equilibrium does not exist. In our case, however, the prize is a function $V(\cdot)$ of contribution and $V(\eps)$ can be so small that players lose the incentive to deviate from all-zero bids. Therefore, we impose $\Pr(b_i)=0$ if $b_i=0, \forall i$ to reinstate this incentive. Indeed, we will show later in \pref{thm:IR} that our mechanism ensures that a player will receive strictly positive payoff if he exerts non-zero effort.}

Another crucial component of our contest model is a prize function $V(\xi_w)$ that we specifically introduce in the contest, which is a (monetary) prize of a common value that is dependent on the (unknown) winner's contribution $\xi_w$. Accordingly, a player $i$ will receive an expected income of $\Pr(b_i) V(\xi_i)$. This function $V(\cdot)$ is common knowledge to all players (e.g., via announcement by the crowdsourcer). 

Each player is characterized by his {\em type}---his marginal cost of exerting effort---denoted by $c_i\in[\uline c, \ovl c]$, where $0<\uline c<\ovl c$. That is, exerting effort $b_i$ will incur a cost of $c_i b_i$ to player $i$. Thus, if the effort profile of all the players is $\vect b:=(b_1,b_2,...,b_n)$, the payoff of player $i$ can be expressed in the following quasi-linear form:
\begin{align*}
\Pr(b_i) V(\xi_i) - c_i b_i.
\end{align*}
Since $\xi_i=g(b_i)$, a player's (ex post) payoff given the contribution strategy profile of all the players $\vect \xi:=(\xi_1,\xi_2,...,\xi_n)$, is
\begin{align}\label{eq:utdef-expost-inv}
\tilde u_i(c_i, \vect \xi) = \frac{\xi_i}{\sum_{j=1}^n \xi_j} V(\xi_i) - h(\xi_i) c_i
\end{align}
where $h:=g^{-1}$ is the inverse function of $g(\cdot)$. We note that $g^{-1}$ exists because $g(\cdot)$ is strictly monotone.

As crowdsourcing typically involves an undefined group of people,
we assume the {\em interim} stage which corresponds to an {\em incomplete-information} setting: each player $i$ is informed of his own type $c_i$ but not of others', yet it is common knowledge that all the $c_i$ are independently drawn from a continuum $[\uline c, \ovl c]$ according to a c.d.f. $F(c)$ or p.d.f. $f(c)=F'(c)$.\footnote{In practice, such a distribution $F(c)$ can be obtained (and published) by the crowdsourcer based on historic data, or---when historic data is not available---assume uniform distribution as widely used in the Bayesian game literature (our model constitutes a Bayesian game with private information being player types and common prior being the type distribution).}  On the contrary, the {\em ex ante} stage corresponds to a no-information setting where players do not know anyone's type including their own, and the {\em ex post} stage corresponds to a complete-information setting where all the players' types are common knowledge. 

The crowdsourcer collects revenue from the aggregate contribution of all the players, and bears the cost of paying for the (variable) prize. The profit cum utility of the crowdsourcer is thus
\begin{align}\label{eq:orgut-def}
\tilde \pi = \nu \sum_{i=1}^n \xi_i - V(\xi_w)
\end{align}
where $\nu$ is the crowdsourcer's valuation of per unit user contribution. This parameter $\nu$ does not appear in prior work where a unity value is always implicitly assumed. However, explicitly modeling this parameter not only homogenizes the dimension of the expression \eqref{eq:orgut-def}, but also allows us to investigate the impact of $\nu$ on key metrics such as the player strategy, prize, profit, and social welfare, which turns out (cf. \sref{sec:orgvalue}) to be an interesting combination of both linearity and nonlinearity.

\section{Proposed Optimal Tullock Contest}\label{sec:analysis}

The solution concept of a game with incomplete information is a pure strategy {\em Bayesian Nash equilibrium}, in which each player plays a strategy that maximizes his expected utility given his belief about other players' types and that other players also play their respective equilibrium strategies. Formally, it is a strategy profile $\vect \xi^{BNE}=(\xi_1^{BNE},\xi_2^{BNE},...,\xi_n^{BNE})$ that satisfies
\[ u_i(c_i,\xi_i^{BNE};\xi_{-i}^{BNE}) \ge u_i(c_i,\xi_i;\xi_{-i}^{BNE}), \;\; \forall \xi_i, \forall i, \]
where $u_i$ is the expected utility of player $i$, defined as
\begin{align}\label{eq:utdef}
u_i(c_i,\xi_i) := \mathbbm E_{\xi_{-i}}[\tilde u_i(c_i, \vect \xi)]
\end{align}
where $\vect \xi=(\xi_i,\xi_{-i})$, and $\tilde u_i$ is defined in \eqref{eq:utdef-expost-inv}. 

Definition \eqref{eq:utdef} can be expanded as follows. In a Bayesian Nash equilibrium, each player's strategy $\xi_i$ is a function of his own type $c_i$ and the common {\em prior}, i.e., each player's belief about all the other players' types. As our setting is symmetric, in that the prior is a common distribution $F(\cdot)$ for all the players,\footnote{In an asymmetric setting, player types follow their respective and generally different distributions $F_i(c_i)$, which is common knowledge. Not only is solving such asymmetric equilibria still an open problem with no analytical solution in general\cite{Konrad09book}, but this setting also makes all players {\em onymous} and thus may engender privacy concerns in practice. \metacom{check Einy13: contest/asym/tullock-contest-w-asym-info.pdf}} we focus on symmetric equilibria in which any player $i$'s equilibrium strategy is specified by a function $\beta:[\uline c,\ovl c]\rightarrow \mathbbm R_+$ as $\xi_i=\beta(c_i), \forall i$.
Therefore, definition \eqref{eq:utdef} can be rewritten based on \eqref{eq:utdef-expost-inv}, as
\begin{align}\label{eq:ut}
u(c,\xi) = p(\xi) V(\xi)  - h(\xi) c
\end{align}
for an arbitrary type $c$ and strategy $\xi$, where 
\begin{align}\label{eq:winprob}
p(\xi) := \int_{\Theta^{n-1}} \frac{\xi}{\xi + \sum_{j=1}^{n-1} \beta(\tilde c_j)} 
								\prod_{j=1}^{n-1} \opd F(\tilde c_j)
\end{align}
in which $\Theta := [\uline c, \ovl c]$. 

Thus, for a particular player $i$, his expected utility is $u_i=u(c_i,\xi_i)$ which can be computed from \eqref{eq:ut}.

\begin{prop}[Existence and Uniqueness of Equilibrium]\label{thm:exists}
Our Tullock contest model admits a unique, monotone decreasing, pure-strategy Bayesian Nash equilibrium.
\end{prop}
Due to space constraint, we defer all the proofs of this paper to \cite{infocom15appendix}.

Henceforth, we will exclusively deal with the equilibrium, and thereby drop the superscript BNE for brevity.

\begin{lem}[Equilibrium Contribution Strategy]\label{lem:strategy}
Given an arbitrary prize function $V(\cdot)$, the (symmetric) equilibrium strategy $\beta(\cdot)$ of our Tullock contest, as in $\xi=\beta(c)$, is implicitly given by
\begin{align}\label{eq:strategy}
p(\xi) V(\xi) - h(\xi) c = \int_{c}^{\ovl c} h(\beta(\tilde c)) \opd \tilde c,\;\;
\forall c\in[\uline c,\ovl c].
\end{align}
\end{lem}
We remark on the following:
\begin{itemize}[leftmargin=0.8em]
\item Equation \eqref{eq:strategy} has an intuitive interpretation: player $i$'s expected utility, as represented by the l.h.s., is determined by his cost advantage relative to the highest-cost player, modulated by his contribution level.
\item There is no closed-form solution to \eqref{eq:strategy} for an arbitrary $V(\cdot)$. In fact, even the fixed-prize case (as in conventional contests) does not have a closed-form solution in general either\cite{Konrad09book}. However, an interesting and counter-intuitive finding arises from maximizing the crowdsourcer's utility through an optimal prize function, on which we remark following the next theorem.
\end{itemize}

\begin{thm}[Optimal Prize Function, Strategy, and Maximum Profit]\label{thm:optimal}
The optimal prize function that maximizes the crowdsourcer's utility (profit) in our Tullock contest is given by
\begin{align}\label{eq:opt-prize}
V^*(\xi_w) = \left[ \beta^{-1}(\xi_w) h(\xi_w)
- \int_{\uline \xi}^{\xi_w} h(\tilde \xi) \opd \beta^{-1}(\tilde \xi) \right]
\Big/ p(\xi_w)
\end{align}
where $\uline\xi=\beta(\ovl c)$ and $\beta^{-1}(\cdot)$ is the inverse function of the equilibrium strategy $\beta(\cdot)$ which, as in $\xi=\beta(c)$, is given by\metafoot{Do not try to simplify by integrating both sides hoping to remove derivative on lhs; doing so will only complicate the expression because you have to integrate rhs w.r.t. $\xi$ not $c$. Equivalently, can write in terms of optimal player effort $b_i = h(\xi_i)$.}
\begin{align}\label{eq:opt-strategy}
h'(\xi) = \frac{\nu}{c + \frac{F(c)}{f(c)}},\; \forall c\in[\uline c,\ovl c].
\end{align}
The induced maximum profit of the crowdsourcer is
\begin{multline}\label{eq:profit}
\pi^* = n \int_{\uline c}^{\ovl c} \bigg[ 
\nu \beta(c) - h(\beta(c)) c \\
+ \frac{F(c)}{f(c)} [ h(\beta(\ovl c)) - h(\beta(c)) ] 	 \bigg] \opd F(c).
\end{multline}
\end{thm}
We will illustrate how to put \thmref{thm:optimal} to use, in \sref{sec:numeric} with a case study. Here we remark on the following:
\begin{itemize}[leftmargin=0.8em]
\item The strategy $\uline\xi$ or $\beta(\ovl c)$, as of the highest-cost or weakest player, is always 0 in all-pay auctions under standard assumptions. However, in Tullock contests, this is not necessarily the case (unless $\ovl c=\infty$), which will be evidenced in \sref{sec:numeric}. The reason is that any Tullock contestant has a positive winning probability as long as he exerts nonzero effort, whereas all-pay auctions perfectly discriminate the weakest bidders who have no chance to win.
\item {\bf Analytical tractability and Practical implication}: An interesting and somewhat surprising observation is that, while ``functionizing'' the contest prize would, intuitively, seem to introduce complexity to conventional, fixed-prize contests, the equilibrium strategy \eqref{eq:opt-strategy} turns out to be much simpler compared to the fixed-prize case (cf. \eqref{eq:strategy-fixed} in \sref{sec:opt-fixed}). In fact, for most and common functions $h(\cdot)$, it \eqref{eq:opt-strategy} can be expressed in closed form. This convenient analytical tractability is in stark contrast to prior art (see \cite{Fey08,Ryvkin10,Wasser13} and a survey \cite{Konrad09book}) where equilibria do not have analytical solutions in general and can only resort to numerical methods.  Theoretically, this lends us a lot of convenience in subsequent technical treatments and other possible future extensions. Practically, this fosters the application of our mechanism via easily-implementable software deployed in web agents and smartphone apps that act on each user's behalf to determine his contribution strategy in a fully distributed manner.\footnote{Note that the optimal prize function \eqref{eq:opt-prize} and maximum profit \eqref{eq:profit} are computed on the other hand by a {\em centralized}, and computationally powerful server, and the computation takes the already-solved $\beta(\cdot)$ \eqref{eq:opt-strategy} as input, unlike in \eqref{eq:strategy-fixed} $\beta_0(\cdot)$ is unknown. In fact, in many cases such as demonstrated in \sref{sec:numeric}, the computation in our mechanism is fairly straightforward.}
\item {\bf Agnosticism of strategy and Practical implication}: Another surprising observation is that the equilibrium strategy \eqref{eq:opt-strategy} is {\em agnostic} to $n$. This is counter-intuitive and in direct contrast with prior findings (again see \cite{Fey08,Ryvkin10,Wasser13,Konrad09book}; also cf. \eqref{eq:strategy-fixed}) where players are {\em antagonistic} to an increasing number of rivals: when the number of players increases, each individual player's chance of winning will be diluted, and hence if the prize is fixed, each player will have to expect a lower utility,
resulting in a {\em disincentive} to participate.  However, now that players can stay agnostic to the participant pool size,\footnote{The agnosticism is achieved by the prize function \eqref{eq:opt-prize} which absorbs $n$ and thereby isolates the player strategy from this number.} they need not worry about an increasing number of rivals, which certainly strengthens the motivation to participate or stay in the campaign. In more practical terms, participants would even not be averse to spreading the awareness and publicity of a campaign, which helps further expand the participant pool.
\end{itemize}

Now we state an important condition pertaining to general incentive mechanisms: {\em individual rationality} (IR) \cite{Jackson03md}. It means that any participating player should receive in equilibrium a nonnegative expected utility, or in other words, each player should be better off or at least remain neutral by participating. In the following, we prove that our mechanism possesses a stronger version of IR.
\begin{prop}[Strict Individual Rationality]\label{thm:IR}
Our Tullock contest with the prize function given by \thmref{thm:optimal} satisfies {\em strict individual rationality} (SIR), where all the players receive strictly positive expected utility, except that a player of type $\ovl c$---which happens with probability zero---expects a surplus of zero (and hence is indifferent in participating).
\end{prop}

Another often-discussed property in mechanism design is {\em incentive compatibility} (IC) \cite{Jackson03md} or truthfulness, which means that all players report their types truthfully. This property is technically irrelevant to our mechanism because, unlike some other mechanisms such as \cite{yang12mobicom,kout13infocom} which determine workers' wages based on worker-reported types (costs or desired payments), our mechanism determines players' reward based on {\em observable} user contributions rather than {\em unobservable} (and private) player types (costs). On the other hand, those other mechanisms can satisfy {\em IR} trivially by paying a wage no less than a worker's reported cost or payment, provided that IC is satisfied; but in our case, satisfying IR requires a carefully designed prize function \eqref{eq:opt-prize} (which is demonstrated by the proof of \pref{thm:IR}).

\section{Optimal Fixed-prize Tullock Contests}\label{sec:opt-fixed}

This section constructs a conventional, i.e., fixed-prize, Tullock contest for the sake of comparison with our mechanism (later in \sref{sec:numeric}).\footnote{It is sound and fair to compare our mechanism with a conventional, fixed-prize counterpart, as our prize function is an extension of the conventional (single) fixed prize. This is also in accordance with \cite{optSymTullock12} which proves that it is optimal to allocate a single (fixed) prize as compared to multiple (fixed) prizes.} However, we augment the prior art of dealing with conventional contests\cite{Fey08,Ryvkin10,Wasser13}, from {\em solving} the equilibrium to {\em optimizing} (as well as solving) the equilibrium, in order to create the most superior benchmark to challenge our mechanism. 

Specifically, we set to find an optimal fixed prize $V_0^*$ that maximizes the crowdsourcer's utility $\pi_0$ through a particular equilibrium $\vect\xi_0^*$, and such a solution needs to be found for every possible value of $\nu$, the valuation of contribution. Formally, the problem is formulated as $\max_{V_0\ge 0} \pi_0$ where
\begin{align}\label{eq:profit-fixed}
\pi_0 = \nu \mathbbm E_{\vect c} \left[ \sum_{i=1}^n \xi_{0 i} \right] - V_0
= n\nu \int_{\uline c}^{\ovl c} \beta_0(c) \opd F(c) - V_0. 
\end{align}
The equilibrium strategy $\xi_0=\beta_0(c)$ is given by \pref{thm:strategy-fix}.
\begin{prop}\label{thm:strategy-fix}
In a Tullock contest with fixed prize $V_0$, the equilibrium strategy $\xi_0=\beta_0(c)$ is implicitly determined by
\begin{align}\label{eq:strategy-fixed}
\int_{\Theta^{n-1}} \frac{\sum_{j=1}^{n-1} \beta_0(\tilde c_j)}
	{[\beta_0(c) + \sum_{j=1}^{n-1} \beta_0(\tilde c_j)]^2 } \prod_{j=1}^{n-1} \opd F(\tilde c_j)
= h'(\xi_0) \frac{c}{V_0}.
\end{align}
\end{prop}

Unfortunately, \eqref{eq:strategy-fixed} does not have an analytical solution. The special case of $V_0=1$ was numerically tackled by \cite{Fey08,Ryvkin10,Wasser13}. In our case, $V_0$ is not given and we need to find the optimal $V_0$ that maximizes $\pi_0$ \eqref{eq:profit-fixed}. In the meantime, $\pi_0$ contains $\beta_0(c)$ \eqref{eq:strategy-fixed} which is analytically intractable. Moreover, the optimal solution $V_0^*$ must not be a value but a function of $\nu$. This problem is technically challenging.

Our solution was inspired by solving the {\em Fredholm equations} using a numerical method described in \cite{numeric90NASA} (Chap. 5). Consider a two-player case for simplicity. The first key idea in our solution is to transform the integral in \eqref{eq:strategy-fixed} into a quadrature sum, by supposing that $V_0$ is given:
\begin{align}\label{eq:strategy-num1}
\sum_{j=1}^m \frac{\beta_0(t_j) f(t_j) \Delta_j} {[\beta_0(c) + \beta_0(t_j)]^2} + R_m(c)
= h_{\beta_0}'(\beta_0(c)) \frac{c}{V_0},
\end{align}
where $t_j, j=1,2,...,m$, are the quadrature points distributed in $\Theta$, $\Delta_j$ is determined by the chosen quadrature scheme (e.g., Gaussian), and $R_m(c)$ is the residual error due to transforming the original integral into the quadrature sum. In our case, a uniform quadrature scheme suffices and thereby $\Delta_j=\Delta_m:=(\ovl c - \uline c)/m$. In addition, since the integrand is atomless ($\beta_0(t_j)=0$ for all $j$ is not an equilibrium because a player will have incentive to deviate by an infinitesimal amount to gain positive utility), the integral can be closely approximated by the quadrature sum for a sufficiently large $m$, in which case $R_m(c)$ can be safely ignored.

The second key idea is to note that, since \eqref{eq:strategy-num1} holds for all $c\in[\uline c,\ovl c]$, it must also hold for all the $c_i$ that equal to the quadrature points $t_j, j=1,2,...,m$. Thus, \eqref{eq:strategy-num1} is further transformed into a system of $m$ nonlinear equations:
\begin{align}\label{eq:strategy-num2}
\Delta_m \sum_{j=1}^m 
\frac{\beta_0(c_j) f(c_j)} {[\beta_0(c_i) + \beta_0(c_j)]^2}
- h_{\beta_0}'(\beta_0(c_i)) \frac{c_i}{V_0} = 0,\; i=1,...,m.
\end{align}
This system can be solved using the Matlab function {\it fsolve}.

Similarly but on a much simpler scale, the profit \eqref{eq:profit-fixed} can be approximated by
\begin{align}\label{eq:profit-num}
\hat\pi_0(\nu, V_0,\beta_0(\vect c)) = n \nu \Delta_m \sum_{j=1}^m \beta_0(c_j) f(c_j) - V_0
\end{align}
where $\beta_0(c_i)|_{i=1}^m$ are the solutions to the system \eqref{eq:strategy-num2}.

The entire solution is outlined in a self-explanatory fashion by the psuedo-code in \aref{alg:opt-fixed}. In the actual implementation, we also added a testing condition to ensure the range $[\uline{V_0},\ovl{V_0}]$ to be large enough to include the peak point of $\hat\pi_0$ (which can be shown to be concave in $V_0$); we also improved the efficiency by adding a stopping condition to terminate the inner loop faster. These are peripheral and hence omitted in \aref{alg:opt-fixed}.
\begin{algorithm}
\caption{Optimizing fixed-prize Tullock contest}\label{alg:opt-fixed}
\KwIn{bounds $\uline c,\ovl c,\uline\nu,\ovl\nu,\uline{V_0},\ovl{V_0}$; $m$; step sizes $\delta_1,\delta_2$}
\KwOut{$\vect V_0^*: \vect\nu\rightarrow\mathbbm R_+$  (a vector of optimal prizes)\\
$\qquad\qquad\vect\pi_0^*: \vect\nu\rightarrow\mathbbm R_+$ (a vector of maximum profits)\\
$\qquad\qquad\vect\Xi_0^*: \vect\nu\rightarrow\mathbbm R_+^m$ (a matrix of which each row is an $m$-element equilibrium strategy profile corresponding to a $\nu\in\vect\nu$)}
$\Delta_m \gets (\ovl c - \uline c)/m$\;
$\vect c \gets \{\uline c, \uline c+\Delta_m, \uline c+2\Delta_m,..., \uline c+(m-1)\Delta_m$\}\;
$\vect V_0 \gets \{\uline{V_0},\uline{V_0}+\delta_2,\uline{V_0}+2\delta_2,...,\ovl{V_0}\}$\;
Create a $|\vect V_0| \times m$ strategy matrix $\vect\xi^A$\;
\For{$\nu \gets \uline\nu$ \KwTo $\ovl\nu$ \rm{(step size: $\delta_1$)} }  {
	$\vect{\hat\pi}_0 \gets \vect 0$\;
	\For{$i\gets 1$ \KwTo $|\vect V_0|$} {
		create $\vect F_m(\vect c,\Delta_m,\vect V_0[i])=0$ according to Eq. \eqref{eq:strategy-num2}\;
		$\vect\xi_0:=(\xi_{0j}|j=1,2,...,m) \gets fsolve(\vect F_m)$\;\label{line:xi}
		$\vect\xi^A[i]\gets \vect\xi_0$ where $\vect\xi^A[i]$ is the $i$-th row of $\vect\xi^A$\;
		$\vect{\hat\pi}_0[i]\gets \hat\pi_0(\nu,\vect V_0[i],\vect\xi_0)$ calculated using Eq. \eqref{eq:profit-num}\;
	}
	$\vect\pi_0^*(\nu)\gets \max_i(\vect{\hat\pi}_0)$\;
	$\vect V_0^*(\nu)\gets \vect V_0[\argmax_i(\vect{\hat\pi}_0)]$\;\label{line:found}
	$\vect\Xi_0^*(\nu)\gets \vect\xi^A[\argmax_i(\vect{\hat\pi}_0)]$\;
}
\end{algorithm}

\section{Performance Evaluation}\label{sec:numeric}

In this section, we compare for the same crowdsourcing campaign a Tullock contest that employs our design against a Tullock contest that adopts a fixed prize, {\it ceteris paribus}.  The former uses the optimal prize function determined by \thmref{thm:optimal}, which we refer to as {\it Tullock-OPF}, and the latter uses the optimal fixed prize determined by \aref{alg:opt-fixed}, which we refer to as {\it OptBenchmark}. We stress that this benchmark is not designed to favor our proposed mechanism, but honestly follows the well-established standard model and, in fact, is fully optimized.

The performance metrics we evaluate are: 
\begin{enumerate}[label=(\alph*)]
\item the {\em equilibrium contribution strategy} of players, which corresponds to revenue;
\item the {\em prize} that the crowdsourcer provisions, which corresponds to cost;
\item the expected {\em utility} of the crowdsourcer, which corresponds to profit;
\item the {\em social welfare} of the campaign, which is the aggregate utility of all the players at equilibrium and we denote by $U:=\mathbbm E_{\vect c}[\sum_{i=1}^n u_i]$; it measures the total surplus of the community by participating in the crowdsourcing campaign.
\end{enumerate}
The rationale is that a company is typically profit-driven and thus concerned about the metrics (a)--(c), while a government agency or a non-profit organization may focus more on the metric (d).

The primary scenario in this section consists of two players whose marginal contribution costs are independently drawn from a uniform distribution $F(c)=c-1, c\in[1,2]$. Player effort $b$ is converted to user contribution $\xi$ according to $\xi=g(b)=\sqrt b$, and hence $h(\xi):=g^{-1}(\xi)=\xi^2$. Such a setup is common in the literature such as \cite{Fey08,Ryvkin10}.  This scenario is then extend to a $n$-player setting.

\subsection{Tullock-OPF: Analytical Results}
Our mechanism can be solved analytically, and the solving process also illustrates how to put \thmref{thm:optimal} to use. First, the equilibrium contribution strategy can be obtained via \eqref{eq:opt-strategy} as
\begin{align}\label{eq:opf-strategy}
\xi = \beta(c) = \frac{\nu}{4c-2}.
\end{align}
Hence $\beta^{-1}(\xi)=\nu/(4\xi)+ 1/2$, and the numerator of \eqref{eq:opt-prize} equals
\begin{align*}
	\xi^2 (\frac{\nu}{4\xi}+ \inv 2) 
+ \int_{\uline \xi}^{\xi} \tilde\xi^2\cdot \frac{\nu}{4 \tilde\xi^2} \opd \tilde\xi
= \frac{\xi^2}{2} + \frac{\nu \xi}{2} - \frac{\nu^2}{24}
\end{align*}
where $\uline\xi=\beta(\ovl c)=\nu/6$. Using \eqref{eq:winprob}, the denominator of \eqref{eq:opt-prize} equals
\begin{align*}
p(\xi) &= \int_1^2 \frac{\xi}{\xi + \frac{\nu}{4c-2} }\opd c 
= \int_1^2 \left(1 - \frac{\nu}{\xi (4c-2) + \nu} \right) \opd c \\
&= 1 - \frac{\nu}{4\xi} \log(4\xi c +\nu -2\xi)\Big|_1^2
= 1 - \frac{\nu}{4\xi} \log\frac{6\xi +\nu}{2\xi +\nu}.
\end{align*}
Therefore, the optimal prize function \eqref{eq:opt-prize} is obtained as
\begin{align}\label{eq:opf-prize}
V^*(\xi_w) = \frac{\frac{\xi_w^2}{2} + \frac{\nu \xi_w}{2} - \frac{\nu^2}{24}}
	{1 - \frac{\nu}{4\xi_w} \log\frac{6\xi_w +\nu}{2\xi_w +\nu}}.
\end{align}
Finally, the induced maximum profit is obtained via \eqref{eq:profit}:
\begin{align}\label{eq:opf-profit}
\pi^* &= 2 \int_1^2 \!\left[ \frac{\nu^2}{4c-2}\! - \!\frac{\nu^2 c}{(4c-2)^2}
\!+\! (c-1) \Big(\frac{\nu^2}{36}\! -\! \frac{\nu^2}{(4c-2)^2}\Big) \right]\! \opd c \nn\\
&= 2 \nu^2 \int_1^2\! \left( \inv{8c-4} + \frac{c-1}{36} \right)\! \opd c 
= \left( \frac{\log 3}{4} + \inv{36} \right) \nu^2
\end{align}

\subsection{OptBenchmark: Numerical Results}

OptBenchmark can only be numerically solved, using \aref{alg:opt-fixed}, of which the parameters are specified by \tref{tab:param}. 
\begin{table}[ht]
\caption{Parameters for \aref{alg:opt-fixed} (OptBenchmark)}\label{tab:param}
\centering
\begin{tabular}{l | c | c | c | c | c | c | c } \hline\hline \noalign{\vskip 0.5ex}
{\bf Parameter} & $\uline\nu$ & $\ovl\nu$ & $\uline{V_0}$ & $\ovl{V_0}$ & $m$ & $\delta_1$ & $\delta_2$ \\ [0.5ex]\hline \noalign{\vskip 0.5ex}
{\bf Value} & 0.5 & 5 & 0.01 & 5 & 100 & 0.5 & 0.01 \\ \hline\hline
\end{tabular}
\end{table}

\fref{fig:traject} reveals the trajectory of finding the duple $(V_0^*,\pi_0^*)$, i.e., the optimal prize and maximum profit, by evaluating over a range of possible prizes, for each $\nu=1,2,3$. The optimal duple is found at the peak of each curve, and the curves clearly demonstrate the concavity of profit versus prize, which confirms the existence and uniqueness of the optimum. Furthermore, how the optimal duple $(V_0^*,\pi_0^*)$ is affected by $\nu$ is examined by \fref{fig:pp-vs-nu}. Interestingly, prize $V_0^*$ coincides with profit $\pi_0^*$ for all the $\nu$'s. This indicates that the revenue of OptBenchmark is {\em double of the cost} (prize). This also implies that, if in \fref{fig:traject} we draw all the other trajectories (in addition to the in-situ three), the peak points of all the trajectories will all fall onto the same straight line $y=x$.

Another observation in \fref{fig:pp-vs-nu} is that both prize and profit are convexly increasing in $\nu$. We will revisit this nonlinear behavior together with other subsequent observations in \sref{sec:orgvalue}.

\begin{figure*}[tb]
\minipage{0.37\textwidth}
  \fbox{\includegraphics[trim=2.5cm 8.2cm 3.4cm 8.6cm,clip,width=\linewidth]{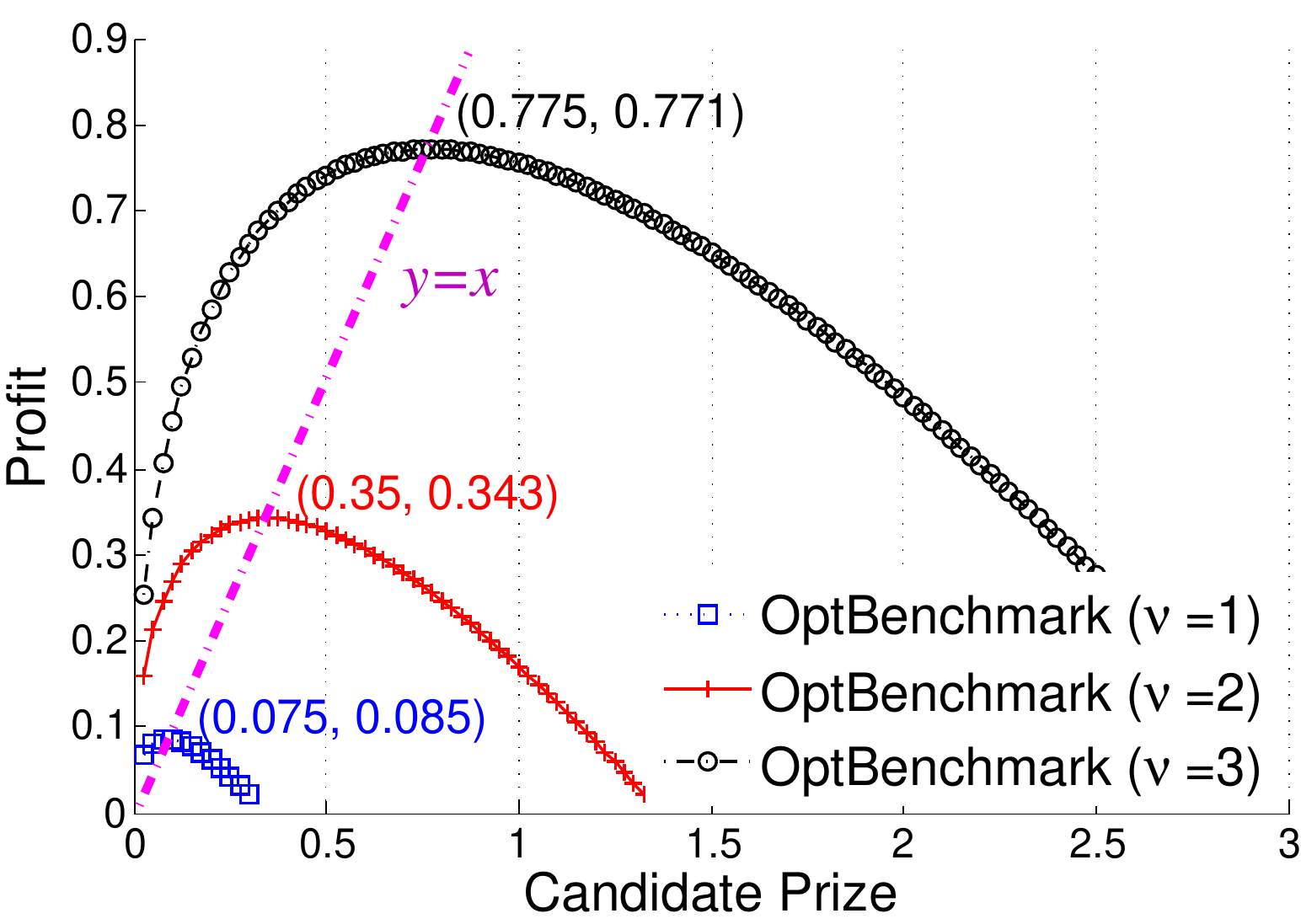}}
  \caption{OptBenchmark: Trajectory of finding the duple of optimal prize and maximum profit, annotated at the peak of each curve.}\label{fig:traject}
\endminipage\hfill
\minipage{0.3\textwidth}
  \fbox{\includegraphics[trim=4.4cm 8.5cm 4.5cm 9cm,clip,width=\linewidth]{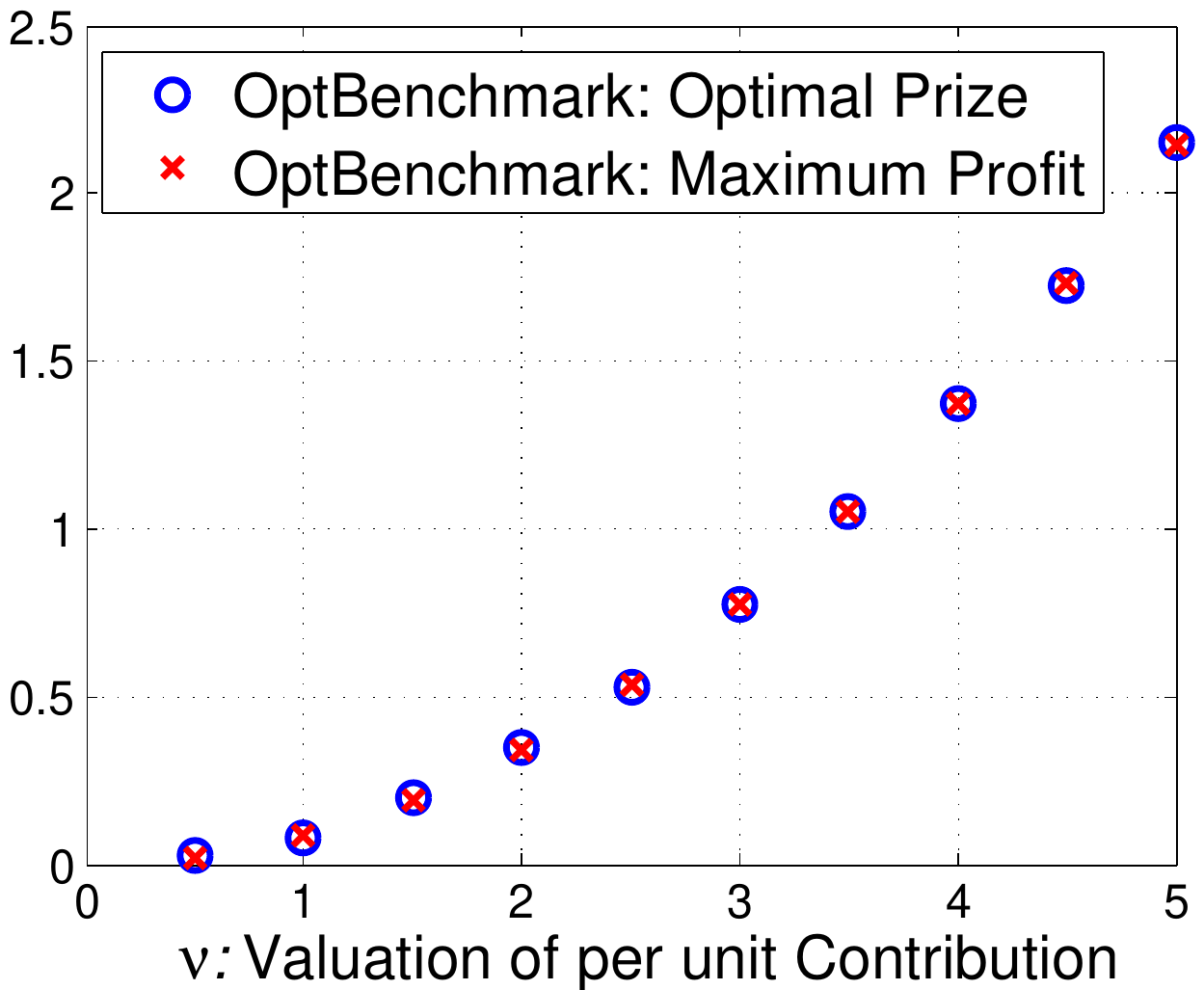}}
  \caption{OptBenchmark: Impact of $\nu$ on prize and profit.}\label{fig:pp-vs-nu}
\endminipage
\setcounter{figure}{3}
\minipage{0.31\textwidth}%
  \fbox{\includegraphics[trim=3.6cm 8.5cm 4.5cm 9cm,clip,width=\linewidth]{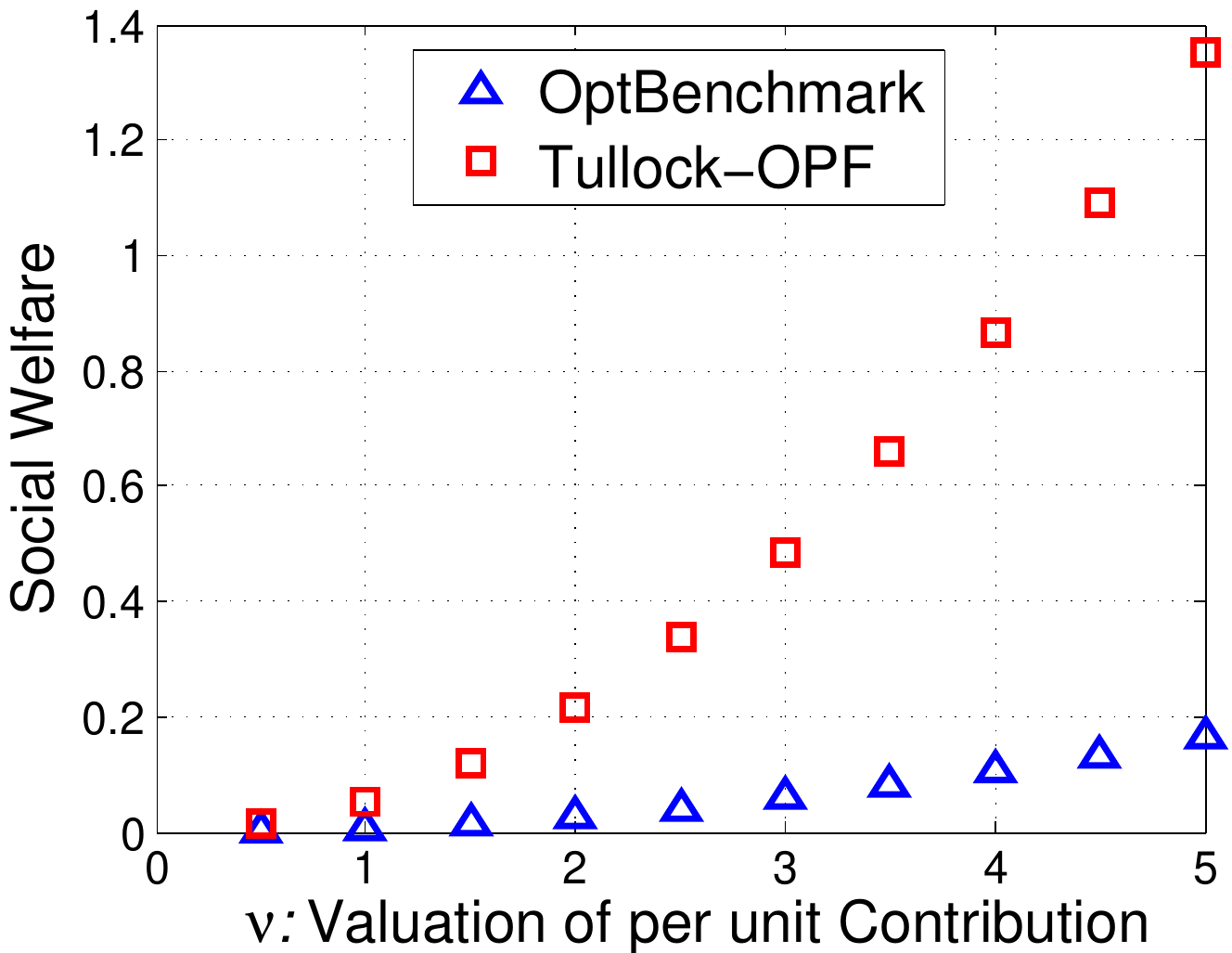}}
  \caption{Comparison of Tullock-OPF against OptBenchmark: Social Welfare.}\label{fig:socwf}
\endminipage
\end{figure*}

\subsection{Comparison}
Given that both Tullock-OPF and OptBenchmark are solved by now, we proceed to compare them with respect to the four metrics mentioned earlier.

\setcounter{figure}{2}
\begin{figure*}[tb]
\centering
\subfloat[Revenue: Equilibrium contribution strategy. Out of the $m$=100 quadrature points, only 20 are shown for better visibility.]{\label{fig:strategy}
\fbox{\includegraphics[trim=2.8cm 6.9cm 3.7cm 7.3cm,clip,width=0.3\linewidth]{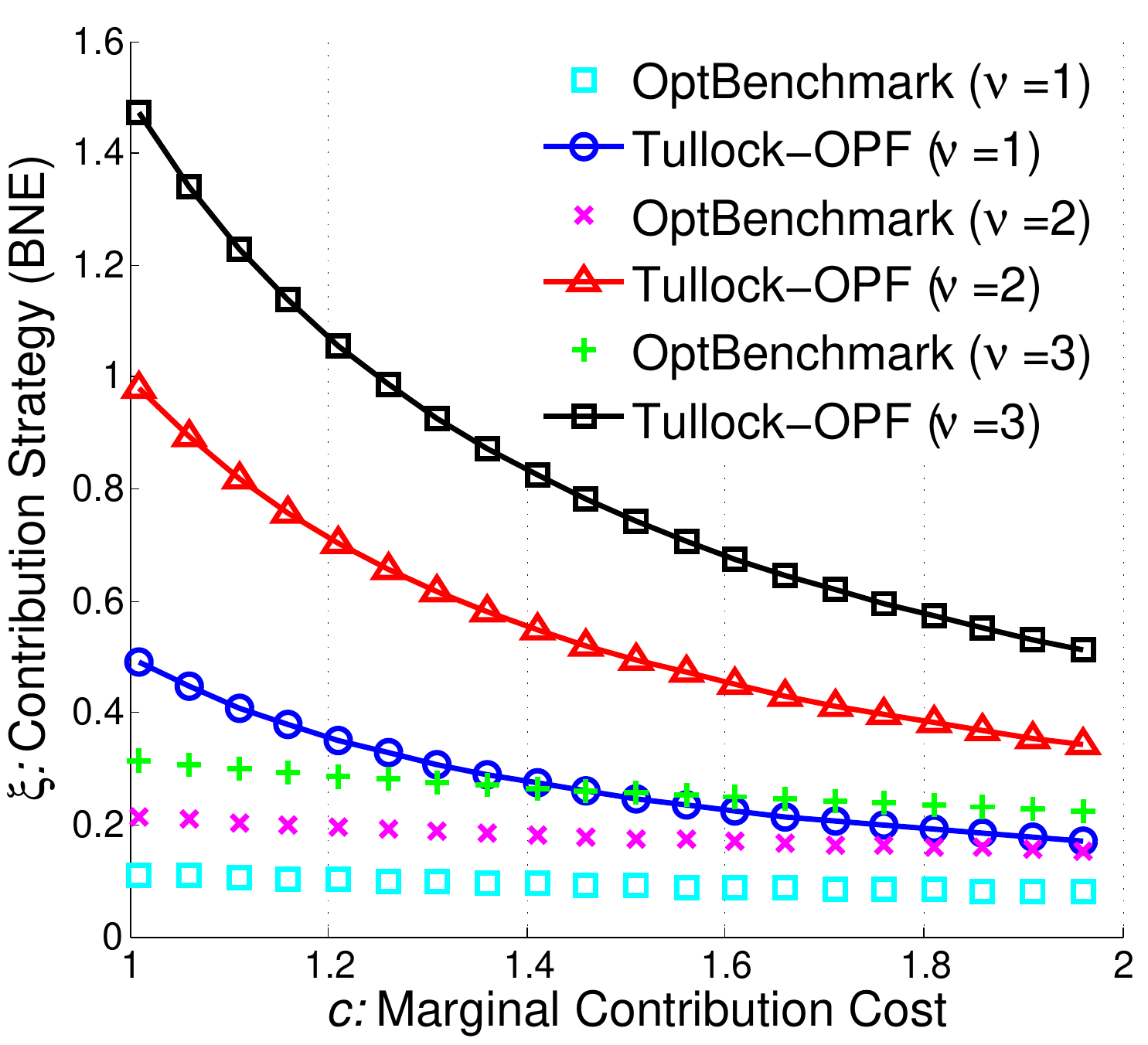}}}\hfil
\subfloat[Cost: Prize. Note that the support of Tullock-OPF is ${\xi\in[\nu/6,\nu/2]}$.]{\label{fig:prize}
\fbox{\includegraphics[trim=3.5cm 8cm 4cm 8.4cm,clip,width=0.32\linewidth]{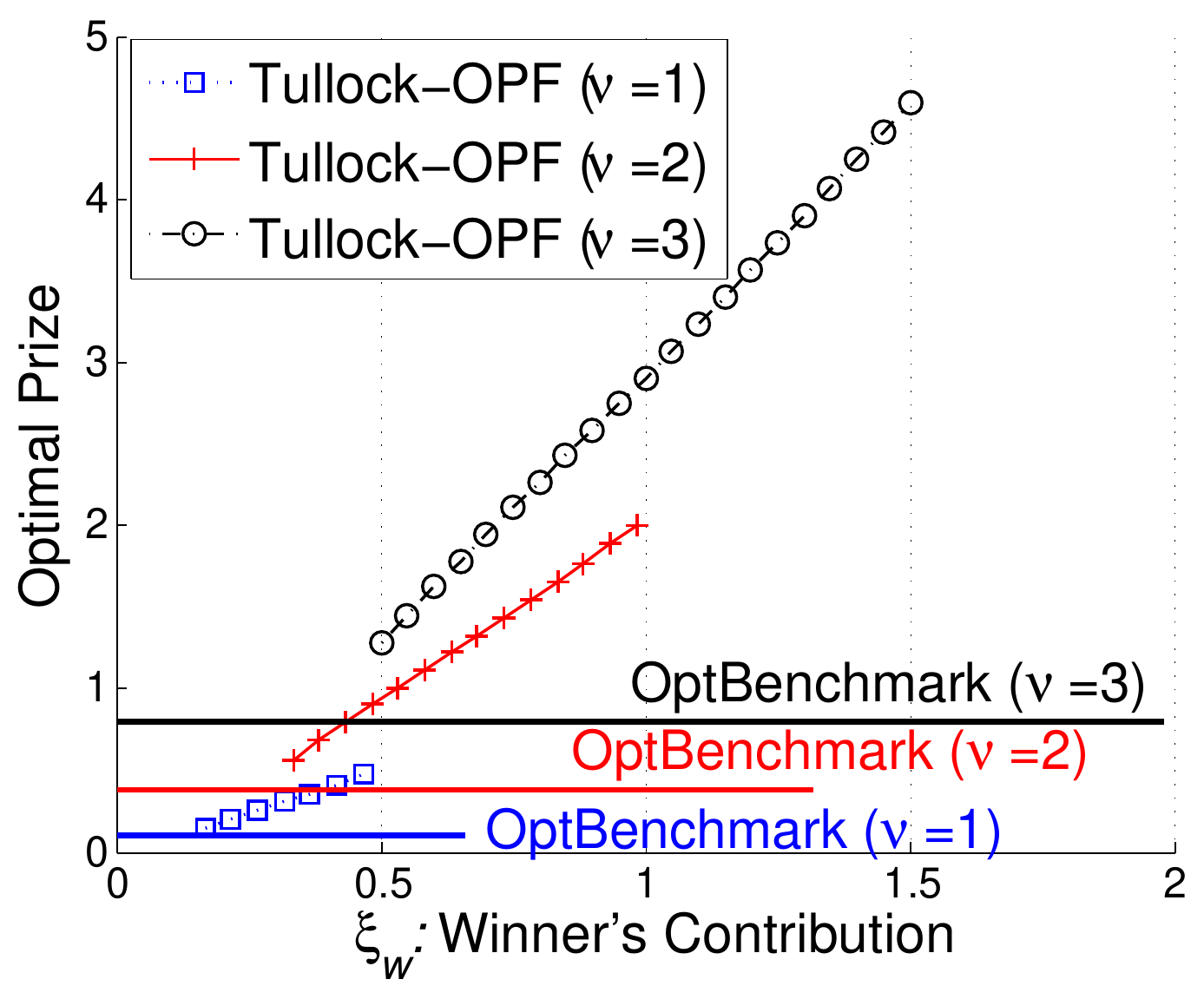}}}\hfil
\subfloat[Profit.]{\label{fig:profit}
\fbox{\includegraphics[trim=4cm 8.4cm 4.5cm 9cm,clip,width=0.325\linewidth]{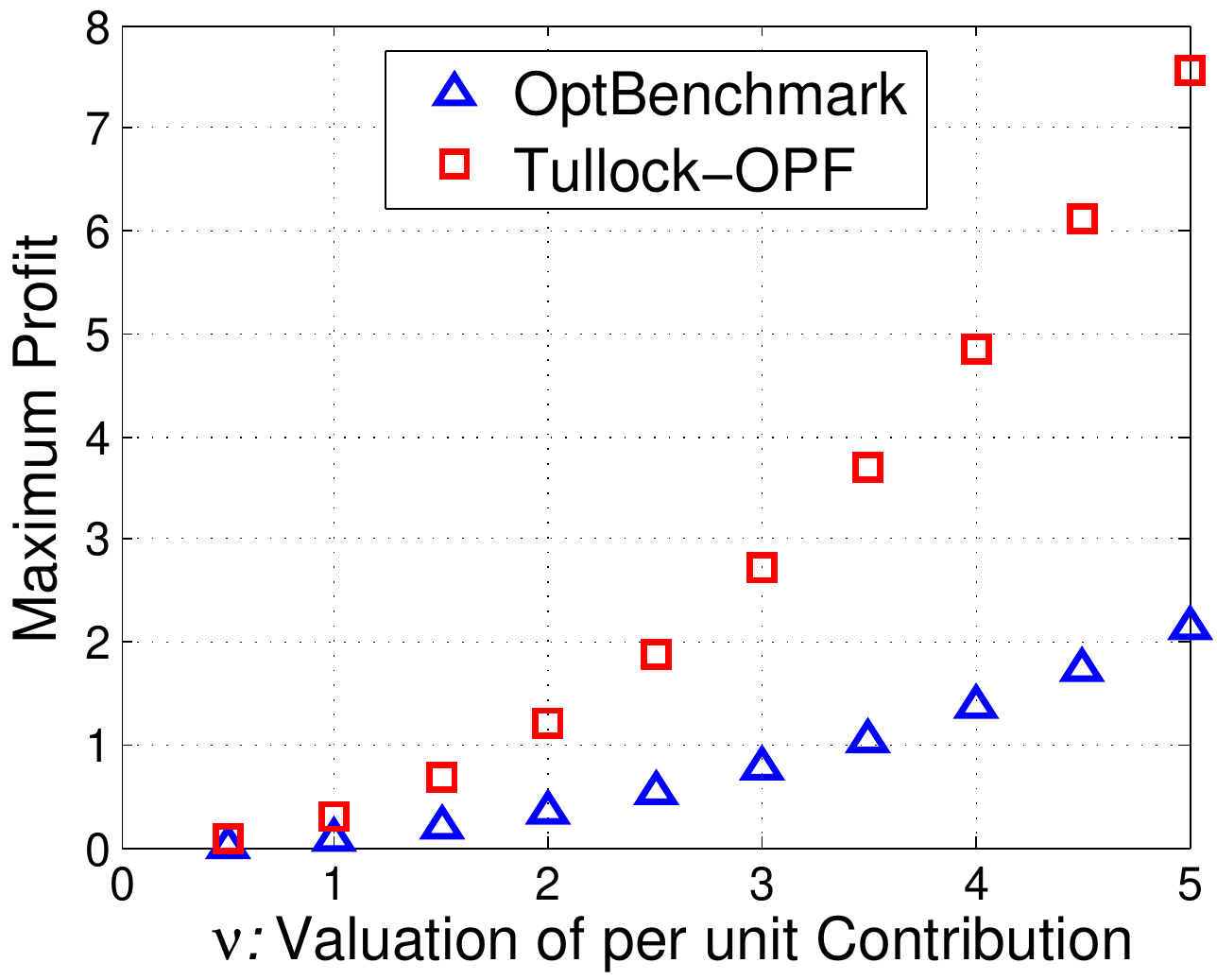}}}
\caption{Comparison of Tullock-OPF against OptBenchmark.}
\label{fig:compare}
\end{figure*}

{\bf Crowdsourcing Revenue}: \fref{fig:strategy} examines the equilibrium contribution strategy of players as a function of player type, for each $\nu=1,2,3$. We remark on three observations. First, in all the cases (3$\times$2 curves), the strategy is monotone decreasing in type, which conforms to our \pref{thm:exists} (which subsumes $V(\cdot)$ being constant). The convex trend is also consistent with the literature: for example, we verified a special case of OptBenchmark with $h(\xi)=\xi$, $\nu=1$, $c\in[0.01,1.01]$, which is the same as \cite{Fey08}, and the result (not reproduced here) exactly matched \cite{Fey08}. Second, for any $\nu$, Tullock-OPF elicits significantly higher contribution than OptBenchmark for every possible type $c$, by about 150\% for high-cost (weak) players and 400\% for low-cost (strong) players. Third, in both mechanisms, the highest-cost or weakest player makes strictly positive contribution, i.e., $\uline\xi=\beta(\ovl c)>0$, indicating a sheer contrast between Tullock contests and auctions where $\uline\xi=0$; here we recall the first remark below \thmref{thm:optimal}.

{\bf Crowdsourcing Cost}: \fref{fig:prize} compares the optimal prize function $V^*(\xi_w)$ in Tullock-OPF against the optimal fixed prize $V_0^*$ in OptBenchmark. Note that the support of $V^*(\xi_w)$ is $[\uline\xi,\ovl\xi]=[\nu/6,\nu/2]$ which follows from \eqref{eq:opf-strategy}. While it is intuitive that each of the three $V^*(\xi_w)$ curves increases in winner's contribution, it is interesting to note that each curve fits a straight line very well, which suggests that the computation of \eqref{eq:opf-prize} could be remarkably simplified in practice via linear approximation. While this advantage should not be overstated as to how it generalizes, it hints at a possible line of future work. Moreover and noteworthily, the diagram shows that the prize offered by Tullock-OPF is generally well above OptBenchmark. This raises an important question that whether this much higher cost to be borne by the crowdsourcer will eventually pay off, which is answered next.

{\bf Crowdsourcing Profit}: \fref{fig:profit} evaluates the maximum profit that a crowdsourcer garners from the two mechanisms. The salient observation is that Tullock-OPF outperforms OptBenchmark by a large profit margin, which strongly corroborates our proposal of using an optimized prize function to supersede the conventional, fixed prize in Tullock contests. For a closer examination, we collate the results into \tref{tab:profit} and calculate the ratio between each pair of profits for all the $\nu$'s. It is interesting to note that the ratio almost remains constant, at about 3.53. This could be explained by the nature of optimization which has pushed the profit to the limit in both mechanisms. In addition, a rigorous analysis as well as investigating to what extent this result can generalize may be worth future exploring.
\begin{table*}[ht]
\caption{Profit Comparison and Ratios}\label{tab:profit}
\centering
\begin{tabular}{l | c | c | c | c | c | c | c | c | c | c } \hline\hline 
{\bf $\nu$} & 0.5 & 1 & 1.5 & 2 & 2.5 & 3 & 3.5 & 4 & 4.5 & 5 \\ \hline 
{\bf OptBenchmark} &  0.0213 & 0.0853 & 0.1927 & 0.3426 & 0.5354 & 0.7710 & 1.0494 & 1.3707 & 1.7347 & 2.1417 \\ \hline
{\bf Tullock-OPF} &  0.0756 & 0.3024 & 0.6805 & 1.2097 & 1.8902 & 2.7219 & 3.7048 & 4.8389 & 6.1242 & 7.5608 \\ \hline
{\bf Ratio} &  3.5533 & 3.5450 & 3.5315 & 3.5307 & 3.5306 & 3.5303 & 3.5303 & 3.5303 & 3.5303 & 3.5303 \\ \hline\hline
\end{tabular}
\end{table*}

{\bf Social Welfare}: In Tullock-OPF, this can be analytically obtained. To solve for $U = \mathbbm E_{\vect c}[\sum_{i=1}^n u_i]$, rather than using the definition \eqref{eq:ut}, we leverage on \lref{lem:strategy} which lends us much more convenience:
\begin{align}
U &= \mathbbm E_{\vect c}\left[\sum_{i=1}^n u_i\right] = n \int_{\uline c}^{\ovl c} \int_{c}^{\ovl c} h(\beta(\tilde c)) \opd \tilde c \opd F(c) \label{eq:socwf}\\
&= n \int_{\uline c}^{\ovl c} \int_{c}^{\ovl c} \frac{\nu^2}{(4\tilde c -2)^2} \opd \tilde c \opd F(c) \nn\\
&= 2 \int_1^2 \frac{\nu^2}{8} \left(\inv{2c-1} - \inv 3\right) \opd c 
= \left(\frac{\log 3}{8} - \inv{12}\right) \nu^2 \nn
\end{align}

One the other hand, the social welfare in OptBenchmark has to resort to numerical methods. To do so, we take the output $\vect\Xi_0^*$ of \aref{alg:opt-fixed}, and denote the row $\vect\Xi_{0}^*(\nu)$ corresponding to each $\nu$ by a strategy profile $\vect\xi_{0}^*$. Thus, we can compute the social welfare as
\begin{align}
U_0 = n \Delta_m \sum_{i=1}^m \left[ f(c_i)
\left( \Delta_m \sum_{j=i}^m {\xi_{0j}^*}^2 \right) \right],
\end{align}
which can be understood by rewriting \eqref{eq:socwf} as
\[ U = n \int_{\uline c}^{\ovl c} f(c) \int_{c}^{\ovl c} \xi^2 \opd \tilde c \opd c \]
and noting that \lref{lem:strategy} applies, without change, to fixed-prize cases where $V(\cdot)$ is a constant.

The results are presented in \fref{fig:socwf}. The observation is exciting: Tullock-OPF outstrips OptBechmark in an even more striking manner as compared to profit in \fref{fig:profit}. The detailed data are collated in \tref{tab:socwf}, which indicates a remarkable improvement of 7--9.3 folds. On the other hand, the nearly constant ratio in \tref{tab:profit} is not duplicated here.

\begin{table*}[ht]
\caption{Social Welfare Comparison and Ratios}\label{tab:socwf}
\centering
\begin{tabular}{l | c | c | c | c | c | c | c | c | c | c } \hline\hline 
{\bf $\nu$} & 0.5 & 1 & 1.5 & 2 & 2.5 & 3 & 3.5 & 4 & 4.5 & 5 \\ \hline 
{\bf OptBenchmark} & 0.0019 & 0.0058 & 0.0155 & 0.0271 & 0.0407 & 0.0600 & 0.0813 & 0.1065 & 0.1336 & 0.1666 \\ \hline
{\bf Tullock-OPF} & 0.0135 & 0.0540 & 0.1215 & 0.2160 & 0.3375 & 0.4859 & 0.6614 & 0.8639 & 1.0934 & 1.3498 \\ \hline
{\bf Ratio} & 6.9693 & 9.2924 & 7.8404 & 7.9649 & 8.2968 & 8.0934 & 8.1308 & 8.1097 & 8.1813 & 8.1038 \\ \hline\hline
\end{tabular}
\end{table*}

It may be puzzling as to why the crowdsourcer can reap higher profit while, at the same time, users altogether also gain higher surplus. This constitutes a ``win-win'' situation which is highly desirable but typically hard to attain. To explain this, we note the following rationales. First, generally speaking, crowdsourcing is not a {\em zero-sum game} like the stock market; rather, it involves a {\em wealth creation} process in which users exert effort to create ``something'' valuable that we have abstracted as ``contribution''. Second, there exists a {\em value asymmetry} between players and the crowdsourcer, where the crowdsourcer typically values contribution higher than players do. This value asymmetry is a common phenomenon in reality: for example in the worldwide emerging Smart City and Smart Nation initiatives nowadays, citizen-generated data such as ambient noise and GPS traces collected by smartphones do not usually bear much value to the phone owners but can be very valuable to a crowdsourcer such as a noise control bureau or a transport company (like Waze). Thus, it makes perfect business sense for a crowdsourcer to provision certain attractive reward to incentivize more user contribution which in turn bears even more value to the crowdsourcer.

\subsection{Impact of Crowdsourcer's Valuation}\label{sec:orgvalue}

Recall that we have introduced a new parameter, $\nu$, to our model in \sref{sec:model} (cf. \eqref{eq:orgut-def}). Herein, we investigate how it affects the various performance indicators.

This effect is explicitly examined with respect to the optimal fixed prize (\fref{fig:pp-vs-nu}), maximum profit (\fref{fig:profit}), and social welfare (\fref{fig:socwf}), where it is clearly shown that the impact of $\nu$ on these three metrics is {\em nonlinear} (convex). Furthermore, this effect is also implicitly examined with respect to the equilibrium contribution strategy (\fref{fig:strategy}) and optimal prize function (\fref{fig:prize}) (in both diagrams one needs to compare across different curves corresponding to different $\nu$'s),
and we see that the impact of $\nu$ on these two metrics is approximately {\em linear}. 

The main message conveyed here is that, if a crowdsourcer increases his valuation of user contribution, e.g. by improving his business processes to better exploit user contribution, his profit and the players' social welfare will both {\em increase faster}. This is an interesting finding uncovered due to our introduction of $\nu$. Moreover, this observation applies to both Tullock-OPF and the conventional case.

Now we dive in deeper to explain the correlation between the above nonlinearity and linearity. On the one hand, the case of OptBenchmark can be nicely explained: (a) the nonlinear profit (\fref{fig:profit}) results from the linear revenue (contribution; \fref{fig:strategy}) and nonlinear cost (prize; \fref{fig:pp-vs-nu}), and (b) the nonlinear social welfare (\fref{fig:socwf}) follows from the nonlinear prize (player's gain; \fref{fig:pp-vs-nu}) and the linear strategy (player's cost; \fref{fig:strategy}).

On the other hand, the case of Tullock-OPF is not that straightforward, since both revenue (contribution) and cost (prize) seem to be linear across different $\nu$'s. In fact, an overlooked fact was that the prize functions in \fref{fig:prize} are {\em shifted} horizontally, and thus one should compare prizes across $\nu$'s for the same {\em winner} rather than for the same amount of {\em contribution}. To do this, a simple way is to compare the maximum (or the minimum) winner contribution $\xi_w$ across different curves, as it can uniquely identify a particular winner. This reveals that the impact of $\nu$ on prize is actually {\em nonlinear}. Combined with the linearity on revenue (\fref{fig:strategy}), this explains why the profit and social welfare in Tullock-OPF are nonlinear in $\nu$. 

\subsection{$n$-Player Case}

In this section, we extend our investigation to $n$ players. We conjecture that the above comparison results will continue to hold in the $n$-player case, and hence we will not repeat the same comparisons. In fact, as Ryvkin \cite{Ryvkin10} pointed out, it is computationally infeasible to numerically solve fixed-prize Tullock contests for an arbitrary large number of players because of the ``curse of dimensionality''. Therefore, this section focuses on Tullock-OPF. In particular, we are interested in how the {\em composition} of a participant pool, i.e., the distribution of the player types, affects the two key metrics, profit and social welfare.

We consider two participant pools: {\it Population-1} draws player types from the same distribution as above, i.e., $F(c)=c-1, c\in[1,2]$, while {\it Population-2} draws from another distribution $G(c)=\frac{c}{2} -\inv 4, c\in[0.5,2.5]$. Thus, Population-2 is more {\em diverse}---or is more {\em uncertain} in player types---than Population-1,
while they statistically share the same mean value (1.5).

Following from \thmref{thm:optimal}, the equilibrium strategy in Population-2 is
$\xi_G = \frac{\nu}{4c - 1}$, and thus the maximum profit is
\begin{align}\label{eq:opf-profit-Gn}
\pi_{n,G}^* \! &=\! \frac{n}{2} \int_{0.5}^{2.5}\! \left[ 
\frac{\nu^2}{4c-1}\! - \!\frac{\nu^2 c}{(4c-1)^2}\! +\! 
(c-\inv 2) \Big(\frac{\nu^2}{81}\! - \!\frac{\nu^2}{(4c-1)^2}\Big) \right]\! \opd c \nn\\
&=\! \frac{n \nu^2}{2} \int_{0.5}^{2.5}\! \left( \inv{8c-2}\! +\! \frac{2c-1}{162} \right) \opd c
= \left( \frac{\log 3}{8} + \frac{2}{81} \right) n \nu^2. \nn 
\end{align}
For Population-1, we leverage \eqref{eq:opf-profit} as a shortcut to obtain
\begin{align}
\pi_{n,F}^* = \left( \frac{\log 3}{8} + \inv{72} \right) n \nu^2.
\end{align}
Social welfare can be solved by referring to \eqref{eq:socwf}:
\begin{align*}
U^*_{n,G} &= \frac{n}{2} \int_{\uline c}^{\ovl c} \int_{c}^{\ovl c} \frac{\nu^2}{(4\tilde c -1)^2} \opd \tilde c \opd c
= \left(\frac{\log 3}{16} - \inv{36}\right) n \nu^2, \\
U^*_{n,F} &= \left(\frac{\log 3}{16} - \inv{24}\right) n \nu^2.
\end{align*}

Thus it immediately follows from the above that
\[  \pi_{n,G}^* > \pi_{n,F}^*, \quad U^*_{n,G} > U^*_{n,F}, \quad \forall n\ge 2, \forall \nu>0.\]
This tells that Population-2 is superior to Population-1 in terms of both profit and social welfare. An insight that may be drawn from this set of results is that population {\em diversity} or {\em uncertainty} is beneficial to Tullock-contest based crowdsourcing for both crowdsourcers and participants.

\section{Conclusion}\label{sec:conc}

To recap, this work has presented a first attempt to use Tullock contests as a new framework to design incentive mechanisms for crowdsourcing. Furthermore, we have explored a novel dimension in the space of optimal Tullock contest design, by superseding the conventional, fixed prize by an optimal prize function for utility maximization. In stark contrast to prior art, we have obtained an analytical solution to the unique Bayesian equilibrium, and found that the equilibrium is robust to an increasing number of rivals. As our model employs a very general contest success function and assumes incomplete information, the mechanism and results would fit a wide range of practical crowdsourcing applications.
For example, WiFiScout\cite{wifiscout14mass} is a mobile app that aims to profile the performance of citywide WiFi access points by eliciting personal experience on WiFi usage from smartphone users. Similarly, OpenSignal\cite{opensignal} aims to construct citywide 3G and 4G LTE cell coverage maps through crowdsourcing too.

The superiority of our design has been demonstrated through extensive evaluations by comparing against a fully-optimized benchmark. Constructing this optimal benchmark significantly extends prior art which only {\em solves} conventional, fixed-prize Tullock contests. This benchmark would be highly relevant to a wider research community for future performance evaluations.

Moreover, we have introduced the crowdsourcer's valuation of user contribution which further extends usual contest models (besides our prize function). It is shown to impact two key metrics---the crowdsourcer's profit and players' social welfare---in a nonlinear (exponential) manner, which bears practical implication on the worth of improving crowdsourcers' business processes. Therefore, this new parameter could be included by future studies (e.g., on radio spectrum auctions or heterogeneous networks) in mathematical models to capture {\em value asymmetry}
and uncover phenomena that are previously unseen.

\metacom{Ultimately, this paper opens up a completely new field of fundamental research in incentive mechanism design for crowdsourcing, using Tullock contests with a functionized prize.}

\bibliography{IEEEabrv}

\end{document}